# Solvation of Silver Ions in Noble Gases He, Ne, Ar, Kr, and Xe

Masoomeh Mahmoodi-Darian,[1] Paul Martini,[2] Lukas Tiefenthaler,[2] Jaroslav Kočišek,[2,3] Paul Scheier,*[,2] Olof Echt *[,2,4]

1   Department of Physics, Karaj Branch, Islamic Azad University, Karaj, Iran
2   Institut für Ionenphysik und Angewandte Physik, Universität Innsbruck, A-6020 Innsbruck, Austria
3   J. Heyrovský Institute of Physical Chemistry of the CAS, 18223 Prague, Czech Republic
4   Department of Physics, University of New Hampshire, Durham NH 03824, USA

Correspondence to:
Paul Scheier; e-mail: paul.scheier@uibk.ac.at, Olof Echt; e-mail: olof.echt@unh.edu

ORCID IDs:
Jaroslav Kocisek    0000-0002-6071-2144
Paul Scheier        0000-0002-7480-6205
Olof Echt           0000-0002-0970-1191

Running title: Solvation of Silver Ions in Noble Gases

Abstract
We use a novel technique to solvate silver cations in small clusters of noble gases. The technique involves formation of large, superfluid helium nanodroplets that are subsequently electron ionized, mass-selected by deflection in an electric field, and doped with silver atoms and noble gases (Ng) in pickup cells. Excess helium is then stripped from the doped nanodroplets by multiple collisions with helium gas at room temperature, producing cluster ions that contain no more than a few dozen noble gas atoms and just a few (or no) silver atoms. Under gentle stripping conditions helium atoms remain attached to the cluster ions, demonstrating their low vibrational temperature. Under harsher stripping conditions some of the heavier noble gas atoms will be evaporated as well, thus enriching stable clusters $Ng_nAg_m^+$ at the expense of less stable ones. This results in local anomalies in the cluster ion abundance which is measured in a high-resolution time-of-flight mass spectrometer. Based on these data we identify specific "magic" sizes $n$ of particularly stable ions. There is no evidence though for enhanced stability of $Ng_2Ag^+$, in contrast to the high stability of $Ng_2Au^+$ that derives from the covalent nature of the bond for heavy noble gases. "Magic" sizes are also identified for $Ag_2^+$ dimer ions complexed with He or Kr. Structural models will be tentatively proposed. A sequence of magic numbers $n$ = 12, 32, 44, indicative of three concentric solvation shells of icosahedral symmetry, is observed for $He_nH_2O^+$.

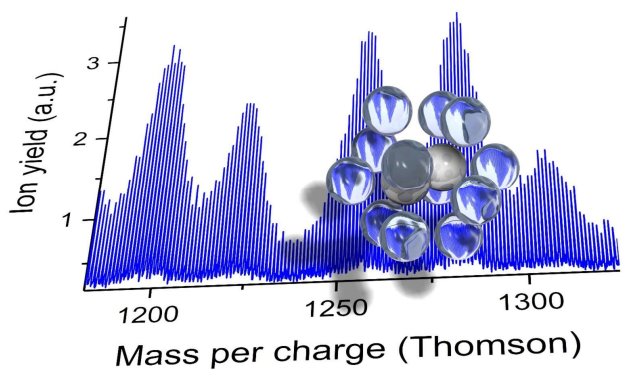

Graphic for Table of Content

## 1. Introduction

Interest in the interaction of coinage metals (M) with noble gases (Ng) has a long history. In cluster science, early attempts to explore the cluster-to-bulk transition involved frozen inert gases doped with silver atoms. Under the right conditions small, neutral silver clusters formed in the matrix; they were examined by optical absorption or Raman spectroscopy.[1-2] Deposition of size-selected clusters has enabled spectroscopy of larger silver clusters.[3-4] Unfortunately, the interaction of the matrix with the embedded clusters is far from negligible; it results in spectral shifts, broadening and line splitting.[5-7]

The matrix approach has been adopted to study clusters of silver and other coinage metals in molecular beams. Duncan and coworkers have applied resonant-enhanced two-photon ionization to measure vibrationally resolved electronic spectra of $NgAg_2^+$ (Ng = Ar, Kr and Xe).[8-9] Diederich et al. have measured the optical absorption spectra of silver clusters embedded in helium nanodroplets (HNDs) by resonant two-photon ionization.[10] Furthermore, the authors measured the spectral shift when the HNDs were doped with neon, argon, krypton or xenon. The interaction with helium is extremely weak but HNDs provide an efficient way to grow neutral clusters and cool them to about 0.37 K.[11-12]

In later work the Rostock group reported optical spectra of silver clusters that were synthesized by passing a supersonic beam of argon clusters through a pickup cell filled with silver vapor.[13] Fielicke et al. have recorded vibrational spectra of $Ag_3$ and $Ag_4$ by subjecting silver clusters tagged with argon atoms to tunable infrared radiation; resonant loss of argon atoms was then detected by softly ionizing $Ar_nAg_m$ with a UV laser near the ionization threshold.[14] The researchers also applied the tagging technique to explore the nature of bonding between argon and mixed gold-silver clusters.[15-16]

In 1995 Pyykö realized that positively charged coinage metal atoms will form very strong chemical bonds with argon, krypton, and xenon.[17] Initial research efforts focused on $XeAu^+$, supposedly the most strongly bound $NgM^+$ dimer, for which Yousef et al. computed a dissociation energy $D_e$ = 1.248 eV at the CCSD(T)/aug-cc-pV5Z level of theory.[18] A mere five years later Seidel and Seppelt reported the synthesis of the first metal-xenon compound with direct gold-xenon bonds.[19]

The strong interaction between $Au^+$ and Xe derives from the covalent character of the bond.[17] The covalency is particularly strong for Xe and Kr but supposedly absent for the weakly ($D_e$ < 0.1 eV[18,20]) bound He and Ne. Ar is considered a borderline case.[21-22]

The bond energies of $NgCu^+$ are weaker than those of the corresponding $NgXe^+$ ion and the covalent contributions are smaller.[17-18,23] For $NgAg^+$ they are even smaller but the bond energy of $AgXe^+$ is still a respectable 0.727 eV.[18] The number of reports devoted to $NgAg^+$ is scarce; even scarcer are reports devoted to the corresponding clusters, $Ng_nAg_m^+$. In addition to the work cited above, theoretical studies of $NgAg^+$ (Ng = He, Ne and Ar) have been published by Tong et al.[24] and Viehland et al..[25] Li et al. have reported ab-initio studies of $Ng_nAg^+$ (Ng = Ar, Kr, Xe, $n \leq 3$)[26-28] and $Ng_nAg^{2+}$ (Ng = Kr, Xe, $n \leq 6$)[29-30] at the CCSD(T)/RPP level of theory. Yasrebi and Jamshidi have used density functional theory (DFT) to investigate the interaction of small neutral and charged silver clusters with a single Ar, Kr or Xe atom and simulate Raman spectra.[5] A comprehensive review of theoretical and experimental work has been published by Grandinetti.[31]

Very few experiments have been devoted to positively charged $Ng_nAg_m^+$, last but not least because their synthesis is challenging. Brock and Duncan have employed resonant-enhanced two-photon ionization to measure electronic excitation of neutral ArAg, KrAg and XeAg.[32] From the vibrationally resolved spectra and ionization energies they derived lower limits on the binding energies of the corresponding $NgAg^+$ cations; the values are consistent with later theoretical work.[18,24] Van der Tol et al. have measured infrared absorption spectra of silver cluster ions tagged with argon.[33] Combined with a DFT study they were able to determine the geometric structure of $Ag_n^+$, $n \leq 13$.

A seemingly straightforward approach to explore the interaction of metal ions with noble gases without involving optical methods is mass spectrometry. Under appropriate conditions one will observe a series of cluster ions $Ng_nM^+$ where the size $n$ covers a large range. Typically, the size dependence of the ion abundance is a smooth function of $n$. Occasionally, however, one observes local anomalies superimposed on the smooth envelope of the ion abundance. These local anomalies may reflect anomalies in the energetic stability of the cluster ions if they form an evaporative ensemble, i.e. if the vibrational excess energy in the ions was at some point large enough to evaporate at least one monomer within the relevant time scale.[34] A useful measure of cluster stability is the dissociation (or evaporation) energy $D_n = E_{n-1} + E_{Ng} - E_n$ where $E_n$ is the energy of $Ng_nM^+$ and $E_{Ng}$ is the energy of an Ng atom. The relation between cluster abundance measured by mass spectrometry and dissociation energy is intricate.[35-36] Qualitatively, however, it is clear that an abrupt decrease of the dissociation energy will cause an enrichment of cluster ion $Ng_nM^+$ at the expense of $Ng_{n+1}M^+$ (provided that evaporation of Ng monomers presents the main dissociation channel).

Recently we reported mass spectra of $Ng_nAu^+$ for Ng = He, Ne, Ar, Kr, Xe.[37] Anomalies in the ion abundance at $Ng_2Au^+$ for Ar, Kr, Xe did, indeed, reflect the large drop in computed dissociation energies that can be traced to covalent bonding in these species. Furthermore, $He_nAu^+$ and $Ne_nAu^+$ showed local maxima at $n = 12$ which are suggestive of solvation shells containing 12 atoms in an icosahedral arrangement. The findings were consistent with expectations based on a simple hard-sphere model proposed by Velegrakis and coworkers that considers the size of the solvated ion relative to that of the ligands.[38]

Mass spectra of silver ions solvated in noble gases have been reported by the Rostock group for $He_nAg^+$, $He_nAg^{2+}$, $Ar_nAg^+$, and $Ar_nAg^{2+}$,[13,39] and by Stace and coworkers for $Ar_nAg^{2+}$.[40] Different approaches were used to form the cluster ions, namely ionization of silver-doped HNDs with a femtosecond laser,[39] and ionization of silver-doped argon clusters either with a YAG pumped dye laser of nanosecond duration,[13] or with 100 eV electrons.[40] A distinct set of abundance anomalies was reported for $He_nAg^+$, namely a local maximum at $He_{12}Ag^+$ and abrupt drops in the abundance beyond $He_{10}Ag^+$, $He_{32}Ag^+$, and $He_{44}Ag^+$.[39] These anomalies were interpreted in terms of two different geometric structures of the solvation shells, one structure containing 10 and 22 atoms in the first and second shell, the other 12 and 32 atoms, respectively. The sequence 12, 32, and 44 has been observed in later work involving ions solvated in either helium or $H_2$, and it is observed in the present work for $H_2O^+$ solvated in helium. The sequence has been attributed to three concentric solvation shells of icosahedral symmetry containing 12, 20 and 12 ligands, respectively.[41-43] The interpretation is consistent with theoretical work.[44-46] $Ar_nAg^+$ and $Ar_nAg^{2+}$ showed no clear abundance anomalies[13,39] while the distribution of $Ar_nAg^{2+}$ suggested enhanced stability for $n = 4$ and 6.[40]

In the present work we adopt a novel approach to form cluster ions $Ng_nAg_m^+$. HNDs are formed in a supersonic expansion and are subsequently ionized by electron collisions. The resulting cations are passed through an electrostatic deflector. The transmitted beam has a specific mass-to-charge ratio $m/z$, hence a specific size-to-charge ratio $N/z$ ($N/z \approx 6 \times 10^4$ is chosen in the present work). The charge state $z$ is broadly distributed depending on the ionization conditions. Accordingly, the size distribution of charged HNDs consists of several slices cut from the original size distribution of neutral HNDs. The ions are then passed through a pickup cell filled with low-density silver vapor and another cell filled with low-density noble gas. The $z$ charge centers in the HNDs act as nucleation centers; a doped HND will carry $z$ singly charged clusters $Ng_nAg_m^+$. Cluster growth releases energy which leads to evaporation of helium but the HNDs are still very large under mild doping conditions. In order to detect them in a subsequent high-resolution time-of-flight mass spectrometer the ions are first passed through a cell containing low-pressure helium at room temperature. Here the HNDs shrink and ultimately undergo fission into singly charged $Ng_nAg_m^+$. Some helium may still be attached to these ions. Careful tuning of the helium pressure in the stripping cell removes most or all of the helium, and eventually some of the heavier, more strongly bound noble gas Ng.

With this approach we observe abundance anomalies that are suggestive of enhanced stability for $He_{12}Ag^+$, $He_6Ag_2^+$ and $He_{12}Ag_2^+$, $Ne_{12}Ag^+$, $Ar_6Ag^+$ and $Ar_{12}Ag^+$, $Kr_4Ag^+$ and $Kr_6Ag^+$, $Kr_{12}Ag_2^+$, and $Xe_4Ag^+$ and $Xe_6Ag^+$. On the other hand, none of the $Ng_2Ag^+$ trimer ions shows evidence of enhanced stability, in contrast to the enhanced ion abundance of $Ng_2Au^+$ previously observed for all noble gases except helium.[37] Geometric structures will be proposed for the "magic" clusters and compared with previous measurements for $Ne_nCu^+$ and $Ar_nCu^+$ by Velegrakis and coworkers,[47] $Ar_nAg^{2+}$ by Stace and coworkers,[48] our measurements of $Ng_nAu^+$,[37] expectations based on a hard-sphere model,[38] and theoretical studies.

## 2. Experiment

The approach adapted for the synthesis and analysis of $Ng_nAg_m^+$ ions is new. Here we focus on the physical processes and implications; technical details will be published elsewhere.[49] The experiment consists of the following steps:

a. Neutral HNDs are produced by supersonic expansion of pre-cooled helium (Messer, purity 99.9999 %, stagnation pressure 20 bar) through a nozzle (5.7 μm diameter, temperature 9.7 K) into vacuum. At these conditions the HNDs will contain an average of about $N = 1.1 \times 10^6$ helium atoms but the size distribution will be broad.[50-51]

b. The expanding beam is skimmed by a 0.5 mm skimmer positioned about 5 mm from the nozzle. Positively charged HNDs are formed by electron ionization (energy 124 eV for HNDs doped with Ar, 70 eV for all other measurements, ionization currents were about 500 μA). Most HNDs experience multiple electron collisions and become $z$-fold charged. Preliminary experiments have shown that the change in droplet size upon ionization is insignificant until the Rayleigh limit is reached at which point small, singly charged helium clusters will be ejected.[52-54] If one stays well below that limit, the number of charges in any specific HND will follow a Poisson distribution. However, the average $z$ increases with droplet size because the ionization cross section scales as $N^{2/3}$, hence a broad HND size distribution implies a distribution of $z$ that is even broader than the Poisson distribution.

c. The charged HNDs (which have not yet been accelerated in an electric field) pass through an electrostatic quadrupole bender which selects ions by their kinetic energy-to-charge ratio. This ratio is proportional to the size-to-charge ratio $N/z$ because the speed distribution of the neutral HNDs is extremely narrow, with a mean speed of about 270 m/s if $T_0 = 9.7$ K.[55] As a result, the transmitted beam is a superposition of narrow slices cut from the size distribution of $He_N^{z+}$, one slice each for $z = 1, 2, 3...$. In the present study, we selected ions with $N/z \approx 6 \times 10^4$.

d. The $N/z$ selected HNDs pass through a pick-up cell containing silver vapor. The average number of captured Ag atoms depends on the droplet size and Ag vapor pressure. For $z > 1$ the charges reside near the surface of the droplet because of their mutual Coulomb repulsion; they will attract the neutral dopant and, for heavy doping, serve as nucleation centers for growth of $Ag_m^+$. Silver has two isotopes of nearly equal abundance. We chose isotopically enriched (99.67 %) $^{109}$Ag (STB Isotope Germany GmbH) in order to reduce ambiguities in the assignment of mass peaks.

e. The ions pass through another cell filled with noble gas at room temperature; pressures ranged from $4 \times 10^{-6}$ to $2 \times 10^{-5}$ mbar (no gas is introduced if silver-helium complexes are the species of interest). Ng atoms colliding with the HND will be captured and attracted to the embedded charges. As a result, $z$ ions $Ng_nAg_m^+$ will be embedded in the HND; $z$, $n$ and $m$ are broadly distributed. Furthermore, the doped HND will have shrunk in size because capture of atoms at and above room temperature and their subsequent aggregation will release a considerable amount of energy: 1 to 2 eV per Ag atom added to the first Ag, and roughly 0.2 eV per Xe atom (less for the lighter noble gases). In the present work we are interested in species with $m \leq 2$ and $n \ll 10^2$, hence no more than a few eV will be released per $Ng_nAg_m^+$. As a rule of thumb, 1 eV of heat released corresponds to the evaporation of about 1600 He atoms.[11]

f. The doped, charged HNDs discussed in the previous section are way too large for conventional mass spectrometry. In order to remove excess helium, the ions are passed through an RF hexapole ion guide filled with helium at room temperature. The helium pressure (ranging from 0.015 to 0.037 Pa in the present work) is adjusted until most or all of the helium has been removed (as evidenced from a mass spectrum, see next paragraph). The length of the cell is 260 mm.

g. In order to measure the size distribution of $Ng_nAg_m^+$ cluster ions produced in the previous steps, the exit of the RF hexapole ion guide is coupled to the entrance ion guide of a commercial time-of-flight mass spectrometer equipped with a double reflectron in W configuration (Micromass Q-TOF Ultima mass spectrometer, Waters). Its resolution measured at FWHM (full-width-at-half maximum) is about $10^4$.

h. Mass spectra are evaluated by means of a custom-designed software that corrects for experimental artifacts such as background signal levels, non-gaussian peak shapes and mass drift over time.[56] The routine takes into account the isotope pattern of all ions that might contribute to a specific mass peak by fitting a simulated spectrum with defined contributions from specific atoms to the measured spectrum in order to retrieve the abundance of ions with a specific stoichiometry. Helium and argon feature one dominant isotope with natural abundances exceeding 99 % but neon, krypton and xenon have three or more naturally occurring isotopes that lead to congested mass spectra.

## 3. Results

Fig. 1 presents sections of mass spectra of charged helium droplets doped with silver and neon, argon, krypton, and xenon (panels b through e, respectively); they show the appearance of $Ng_nAg_m^+$ ions. The spectrum in panel a was measured with no noble gas in the pickup cell, leading to the appearance of $He_nAg_m^+$. The mass ranges were chosen to illustrate some salient features. A comprehensive appraisal of relevant information extracted from these spectra will be presented further below.

The mass spectrum in Fig. 1a is relatively straightforward to decipher because helium is essentially monisotopic, and we used isotopically enriched (99.67 %) $^{109}$Ag. Mass peaks belonging to one of the five most prominent ion series have been marked by symbols and connected by lines, namely $He_n^+$, $He_nH_2O^+$, $He_nAg^+$, $He_nH_2OAg^+$, and $He_nAg_2^+$.

By and large the peak amplitudes of the various cluster ion series vary smoothly with size $n$ but there are several noticeable exceptions (note the logarithmic $y$-scale). The most conspicuous ones are abrupt drops after $He_{12}H_2O^+$ at 66 u and $He_{12}Ag^+$ at 157 u.

Fig. 1b displays a mass spectrum of HNDs doped with neon. The two most prominent ion series, $Ne_n^+$ and $Ne_nAg^+$, have been flagged. Each ion with a specific composition produces several mass peaks because neon has three isotopes, $^{20}$Ne (natural abundance 90.48 %), $^{21}$Ne (0.27 %) and $^{22}$Ne (9.25 %). For each value of $n$ we have marked the most abundant isotopologue, namely isotopically pure $^{20}$Ne for $n < 10$ and isotopologues containing one $^{22}$Ne for $n > 10$. For $n = 10$ both isotopologues have been marked. The Ne$_n$Ag$^+$ series exhibits an abrupt drop at $n = 12$. The Ne$_n^+$ series features a drop at $n = 14$, well known from previous reports that employed other methods for the synthesis of cluster ions.[57]

A mass spectrum of HNDs doped with argon is presented in Fig. 1c. The natural abundance of the main argon isotope, $^{40}$Ar, is 99.60 %, hence several homologous ion series can be easily identified, most prominently Ar$_n^+$ and Ar$_n$Ag$^+$. Anomalies in the Ar$_n$Ag$^+$ series at $n = 6$ and 12 are marked. The deep minimum at $n = 20$ in the Ar$_n^+$ series has been observed before by several authors employing various methods to generate the ions.[58-61] Two weaker ion series, not marked in Fig. 1c, are easily identified by visual inspection; they are due to Ar$_n$H$_2$O$^+$ and Ar$_n$H$_2$OAg$^+$.

Krypton has 6 naturally occurring isotopes, including 4 with abundances above 10 % ($^{82}$Kr at 11.58 %, $^{83}$Kr at 11.49 %, $^{84}$Kr at 57.00 %, and $^{86}$Kr at 17.30 %); its average mass is 83.8 u. Hence the spectrum of Kr doped HNDs in Fig. 1d (note the linear $y$-scale) is congested even though it covers only a small cluster size range. Members of ion series that one would expect based on the previous discussion are readily observed: Kr$_n^+$, Kr$_n$H$_2$O$^+$, Kr$_n$Ag$^+$, and Kr$_n$H$_2$OAg$^+$. The inset in Fig. 1d covers a wider mass range plotted with a logarithmic $y$-scale. For each value of $n$ a group of 3 "bumps" appears beyond $m \approx 600$ u; they are due to Kr$_n^+$, Kr$_n$Ag$^+$, and Kr$_n$H$_2$OAg$^+$ whose isotopologues overlap.

Xenon has an even richer isotope pattern than Kr; 5 of its 9 isotopes have abundances above 8% ($^{129}$Xe at 26.44 %, $^{131}$Xe at 21.18 %, $^{132}$Xe at 26.89 %, $^{134}$Xe at 10.44 %, $^{136}$Xe at 8.87 %, average mass 131.3 u). Here, the overlap between groups of isotopologues of different homologous ion series is even more severe. For example, the average mass of Xe$_3^+$ exceeds that of Xe$_2$H$_2$O$^{109}$Ag$^+$ by only 4.4 u but the isotopologues of Xe$_3^+$ cover about $\pm 10$ u; the FWHM width of the isotopologue distribution is $\approx 10$ u. The inset in Fig. 1e shows that the "bumps" due to the different ion series merge above $\approx 1000$ u.

Peaks heights are poor indicators of ion abundance: they may be affected by mass spectral overlap with other ions, and they are subject to larger statistical fluctuations than the ion abundance that is derived by integrating over mass peaks and summing over all isotopologues. The abundance distributions of various ion series are derived from the mass spectra in Fig. 1 by a custom-designed software that takes into account all possible isotopologues, and contributions from impurities, background, and isotopologues of other ions.[56] Results are compiled in Fig. 2, using a logarithmic ordinate. Error bars (one standard deviation) are indicated; most of them are smaller than the symbol size.

Fig. 2 includes abundance distributions of Ng$_n$Ag$^+$ ion series for all noble gases. Distributions for He, Ne and Ar (panels a, c, d) reveal anomalies at $n = 12$ whereas distributions for Kr$_n$Ag$^+$ and Xe$_n$Ag$^+$ (panels e, g) reveal anomalies at $n = 4$ and 6. Two distributions are shown for He$_n$Ag$^+$, measured with different pressures in the helium stripping cell (0.0149 and 0.0168 Pa, respectively). They demonstrate that the local anomaly at $n = 12$ does not change even though the envelope shifts to the left as the He pressure is increased.

Also shown in Fig. 2 are distributions for He$_n$Ag$_2^+$ and Kr$_n$Ag$_2^+$ (panels b and f, respectively); they feature anomalies at $n = 12$. Other, weaker anomalies are present in some distributions, e.g. at $n = 6$ for He$_n$Ag$_2^+$ and Ar$_n$Ag$^+$.

The abundance distribution of He$_n$H$_2$O$^+$ is presented in Fig. 2h (note the extended size range). These data are interesting because they show a sequence of anomalies at $n = 12, 32, 44$ which has been reported before for some other ions solvated in helium or hydrogen.[41-43] The significance of another anomaly, at $n = 25$, is questionable because of possible interference from contaminants.

We have analyzed the abundance distributions of many other ion series, including Ng$_n$H$_2$O$^+$, Ng$_n$H$_2$OAg$^+$, and Ng$_n$Ag$_2^+$ for all noble gases. Most of them were excluded from Fig. 2 for one of two reasons: i) they are marred by large uncertainties. For example, the masses of Ar$_n$Ag$_2^+$ and Ar$_{n+5}$H$_2$O differ by only 0.013 u. Their mass peaks merge in the relevant mass range, above 218 u, and the abundance of their minor isotopologues (containing one or more $^{36}$Ar) is not sufficient to tell them apart. ii) The distributions do not reveal large deviations from a smooth envelope. Furthermore, distributions of pure noble gas clusters Ng$_n^+$ are not shown because they resemble those recorded by conventional techniques, i.e. by post-ionization of noble gas clusters formed in supersonic beams or by aggregation in neutral HNDs.[57,59,61-62]

As discussed in Section 2, the charged HNDs that exit the pickup cell where they are doped with silver and noble gas atoms are very large, too large for conventional mass spectrometry. Most of the excess helium is removed upon their passage through a stripping cell that is filled with helium; the pressure is adjusted in order to optimize the yield of ions of interest. After boiling off the helium the cluster ions will eventually also evaporate heavier, more strongly bound noble gas atoms because the stripping gas is at room temperature.

Fig. 3 displays a sequence of mass spectra of charged HNDs doped with silver but no additional noble gas, recorded with different pressures in the stripping cell. The two dominant groups of mass peaks are due to He$_n$Ag$^+$ and He$_n$Ag$_2^+$. The envelopes of these groups shift toward lower values of $n$ as the pressure in the stripping cell increases from $p_{He}$ = 0.014 Pa to 0.018 Pa and to 0.0235 Pa (panels a, b and c, respectively). Furthermore, the distributions become narrower as they terminate at $Ag^+$ and $Ag_2^+$. However, anomalies in the ion abundance do not shift although they will change their shape. For example, $He_{12}Ag^+$ and $He_6Ag_2^+$ form local maxima in panel a but ledges in panel b. Either way, the spectra suggest that the dissociation energies of these ions are enhanced compared to the next larger cluster ion.

A few other, weaker ion series emerge in Fig. 3. $He_n^+$ appears in the low mass range of panel a. $He_nH_2OAg^+$ and $He_nH_2OAg_2^+$ appear in panels b and c midway between the main mass peaks that are due to water-free ions. Increasing the stripping cell pressure shifts the envelopes to the left. Eventually, $H_2OAg^+$ at mass 127 u and $H_2OAg_2^+$ at 236 u become the most intense peaks of these series.

It is worth pointing out that we do not observe any ions complexed with $OH^+$ or $H_3O^+$; those ions are commonly observed as products of ion molecule reactions if complexes containing more than one $H_2O$ are ionized.[63] Their absence demonstrates that the pickup of water occurs after ionization, most likely in the He stripping cell where cluster ions undergo a very large number of collisions.

4. Discussion

**Cluster Ion Abundance and Stability.** The main goal of the present work is to identify cluster ions $Ng_nAg_m^+$ of enhanced stability, i.e. ions whose dissociation energy $D_n = E_{n-1} + E_{Ng} - E_n$ is relatively large compared to that of the next larger one. Local anomalies in the size dependence of $D_n$ come in several forms; common ones are a local maximum ($C_{60}$ and $C_{60}^+$ belong to this category),[64-65] or ledges where $D_n$ is approximately constant up to a magic size $n$, and then drops to a lower value which, again, remains approximately constant for the next several values of $n$. Numerous examples exist for this category which is encountered when solvent molecules physically interact with an ion, via dispersion and polarization force. Closure of the first solvation shell will often be accompanied by an abrupt drop of $D_n$. Closure of subsequent shells may also be marked by abrupt decreases of $D_n$.[41-46] Similar anomalies may also arise from subshell closures which come in different forms, for example if an alkali dimer ion is being solvated in helium,[66-67] if the cluster grows facets on top of an underlying ordered structure, as observed for pure noble gas cluster ions,[61,68-69] or if a commensurate layer grows on top of a corrugated surface, e.g. on $C_{60}^+$, $C_{70}^+$, or clusters of fullerenes.[70-71]

The present work is based on mass spectra. The link between the measured abundances of cluster ions and their dissociation energies has been established within the framework of the evaporative ensemble by Klots, and by Hansen and coworkers.[34,36] Under the premise that the ensemble of cluster ions being interrogated has evolved from an initial, statistical ensemble of larger clusters via evaporations, one finds that an anomaly in the stability at size $n^*$ is reflected by an anomaly in the abundance at $n^*$. However, the size dependences of $D_n$ and the ion abundance in the vicinity of $n^*$ may differ considerably in shape.[72] Nevertheless, the correlation between the two quantities becomes semiquantitative for cluster ions whose heat capacity at the relevant experimental temperature becomes much smaller than the classical limit, which applies to the solvation of atomic ions, or molecular ions lacking low-frequency vibrational modes, in helium.[67,70]

The concept of temperature for an ensemble of small, isolated clusters subject to evaporative cooling is subtle because the ensemble is not canonical.[34,73] For our purpose it will suffice to note that the temperature will be a function of $D_n$ and a weak function of the time elapsed between the excitation of the cluster ensemble and the mass analysis. As a rule of thumb, the vibrational temperature of clusters, probed roughly 1 ms after excitation, will settle at 50 to 60 % of their bulk boiling point at 1 atm.[74-75] Note that the dissociation energy of small $Ng_nAg^+$ cluster ions will be larger than the cohesive energy of Ng in bulk form; their temperatures will be higher than predicted by the rule of thumb.

A few caveats are in order. First, the ion abundance provides no clue to *absolute* dissociation energies. Second, dissociation energies are usually taken to refer to zero temperature, i.e. they measure the energy difference between ground state structures. Evaporations, however, require finite temperatures, and one should really speak of free energies which may involve significant contributions from configurational entropies.[76] Third, the current experiments use an unconventional approach to excite cluster ions by heating them in a stripping cell filled with helium at room temperature. In this approach the cluster ions are subject to multiple collisions. For a typical He pressure of 0.02 Pa and an initial droplet size of $N \approx 1.1 \times 10^6$ atoms, corresponding to a diameter of 46 nm, the collision frequency will be approximately $1 \times 10^7$ s$^{-1}$. Each fully inelastic collision will transfer $\approx 0.025$ eV, causing the evaporation of $\approx 40$ He atoms. Therefore the cluster ions do not form an evaporative ensemble which assumes multiple evaporations following a single initial excitation without any re-excitation. Nor do they form a canonical ensemble because they are not truly thermalized in the stripping cell, else they would shed all their noble gas ligands resulting in bare silver cluster ions $Ag_m^+$.

**A Hard-Sphere Model.** With these general remarks we are now prepared to discuss the ramifications of our experimental results. The discussion will be guided by a simple hard-sphere model, proposed by Velegrakis and coworkers,[38] that applies to solvation of atomic ions with non-directional bonding. According to this model, the number $n_1$ of solvent atoms in the first coordination shell, and its structure, will depend on just one parameter, the relative size $\sigma^* = R_{M-Ng}/R_{Ng-Ng}$ where $R_{M-Ng}$ is the distance between the ion and the ligand, and $R_{Ng-Ng}$ is the distance between adjacent ligands. The predictions are summarized in the lower half of Fig. 4: For large values of $\sigma^*$ an icosahedral arrangement with coordination number $n_1 = 12$ will be preferred. Structures with smaller coordination numbers such as a square antiprism (a twisted cube, symmetry $D_{4h}$, $n_1 = 8$), octahedron ($O_h$, $n_1 = 6$), or tetrahedron ($T_d$, $n_1 = 4$) will be preferred as the value of $\sigma^*$ decreases. The relative sizes calculated for the structural transitions within the hard-sphere model are indicated in Fig. 4 below the abscissa. For example, the solvated ion will perfectly fit into an octahedral shell if $\sigma^* = 0.707$. As the ion grows (or the ligands shrink), the ligands will no longer be in direct contact with each other, and the shell converts to a square antiprism which will provide an exact fit for the ion at $\sigma^* = 0.823$. The shell will transform to an icosahedron if $\sigma^*$ exceeds 0.823. The basic predictions of this simple hard-sphere model are consistent with theoretical work that employ more realistic potentials for the ion-ligand and ligand-ligand interaction.[38,77-78] However, other features of the assumed potentials, such as their softness, will then also affect $n_1$.

Values of $\sigma^*$ estimated for $Ng_nAg^+$ are indicated in Fig. 4 above the abscissa. Values of $R_{M-Ng}$ have been taken from the equilibrium distance of $NgAu^+$ calculated by Yousef et al..[18] $R_{Ng-Ng}$ is equated with the nearest-neighbor distance in noble-gas crystals which adopt close-packed structures.[79] For Ne through Xe these crystal values deviate less than 3 % from the well-known equilibrium distance of the diatomics; for He this choice avoids bias arising from the exceedingly large zero-point motion in $He_2$.[80] Note that values of $\sigma^*$ will necessarily be approximations because the actual ion-ligand and ligand-ligand distances will depend on cluster size.

Based on the hard-sphere model one would expect an icosahedral structure for the first solvation shell in $He_nAg^+$ and $Ne_nAg^+$ and an octahedral solvation shell for the heavier noble gases. Prominent anomalies in the abundance distributions (Fig. 2) of $He_nAg^+$ and $Ne_nAg^+$ at $n = 12$, and at $n = 6$ for $Kr_nAg^+$ and $Xe_nAg^+$ are consistent with this prediction. However, $Ar_nAg^+$ has a rather weak anomaly at $n = 6$, a stronger one at $n = 12$ which is not explained by the hard-sphere model, and possibly a few others. Anomalies at $n = 4$ for $Kr_nAg^+$ and $Xe_nAg^+$ are left unexplained as well.

**Previous Work on Silver-Noble Gas Cluster Ions.** Theoretical work published for $Ng_nAg^+$ is scarce, except for some studies of $NgAg^+$ dimer ions[18,24-25,81-82] and the $Ar_2Ag^+$ trimer ion.[82] Dissociation energies computed for $NgAg^+$ by Yousef et al.[18] are summarized in Fig. 5 which also presents dissociation energies of $Ar_nAg^+$, $Kr_nAg^+$, and $Xe_nAg^+$ computed by Li et al. at the CCSD(T) level for $n \leq 3$.[26-28]

Given the absence of any theoretical work on $Ng_nAg^+$ with $n > 3$, we consult work on $Ng_nAg^{2+}$ dications. This choice is not unreasonable because the calculated bond lengths $R_{M-Ng}$ do not change much as the charge state $z$ increases from 1 to 2. The relative sizes $\sigma^*$ decrease from 0.70 to 0.69 for argon,[48] from 0.67 to 0.61 for Kr,[30] and from 0.64 to 0.62 for Xe.[29] Within the hard-sphere model one would still expect octahedral symmetry for their first solvation shells.

Dissociation energies of $Ar_nAg^{2+}$ and $Kr_nAg^{2+}$ dications, calculated for $n \leq 6$, are included in Fig. 5.[30,48] Values calculated for $Xe_nAg^{2+}$ at the CCSD(T) level show a similar trend.[29] These data reveal enhanced stability at $n = 4$ in agreement with our data for Kr and Xe, but not Ar. Interestingly, the computed ground state structures are identical for Ar, Kr, Xe, namely a planar square ($D_{4h}$ symmetry) for $n = 4$, and a distorted octahedron ($D_{4h}$ square bipyramid) for $n = 6$. Theoretical work does not extend beyond $n = 6$, but our observation of anomalies at $n = 6$ suggests that the first solvation shell does indeed close at $n = 6$ for Ar, Kr, Xe. Theoretical work will be needed to reveal the geometric structure of the apparently magic $Ar_{12}Ag^+$.

We briefly turn to previously reported mass spectra of $Ng_nAg^{z+}$. The Rostock group explored ionization of Ag-doped HNDs with a femtosecond laser. They observed steps in the abundance distribution of $He_nAg^+$ at $n = 10, 12, 32$, and 44.[39] The steps at $n = 32$ and 44 are not discernible in our data, see Fig. 1a. Their interpretation was that $Ag^+$ occurs in one of two different electronic states, one dressed with a total of 10 and 32 He atoms upon closure of the first two coordination shells, the other with 12 and 44 He atoms. A more convincing structural model for the magic numbers 12, 32, 44 will be discussed below.

A mass spectrometric study of $Ng_nAg^+$ ions has been reported by Tiggesbäumker and coworkers.[13] They doped neutral argon droplets, formed in an adiabatic expansion, with silver. Subsequent multiphoton ionization in the visible spectrum produced $Ar_nAg^+$ and $Ar_nAg_2^+$ containing approximately 15 to 50 argon atoms. No statistically significant abundance anomalies could be discerned within that range. A singular peak at lower mass was assigned to $Ar_{11}Ag^+$; it defies any convincing explanation.

Stace and coworkers observed $Ar_nAg^{2+}$, $n \leq 8$, by passing a supersonic beam of argon clusters through vapor of silver and subsequent electron ionization of the mixed clusters at 100 eV.[48] The abundance distribution featured a maximum at $n = 4$ and a sharp decline after $n = 6$, in qualitative agreement with the computed dissociation energies. The researchers were solely interested in dications; they did not mention monocations in their report.

**Coinage Metal Ions Solvated in Noble Gases.** For further insight we consult research on $Cu^+$ and $Au^+$ solvated in noble gases. Velegrakis and coworkers have reported mass spectra of $Ne_nCu^+$ and $Ar_nCu^+$, formed by laser vaporization of copper into an expanding gas pulse.[47] They observed an abrupt drop in the abundance after $Ne_{12}Cu^+$, similar to our finding for $Ne_{12}Ag^+$. Electronic structure calculations with DFT revealed, indeed, that the first solvation shell closes at $Ne_{12}Ag^+$ which has icosahedral symmetry; its computed dissociation energy is more than thrice that of $Ne_{13}Ag^+$. Still, the result is surprising because the $NeCu^+$ bond length is considerably shorter than that of $NeAg^+$. As a result, the relative size $\sigma^*$ decreases from 0.84 for $Ne_nAg^+$ to 0.74 for $Ne_nCu^+$, way below the region where the hard-sphere model would predict an icosahedral solvation shell. The relative size shrinks even further if one adopts the $NeCu^+$ bond length computed by Froudakis et al.[47] rather than the bond length computed by Yousef et al..[18]

The $Ar_nCu^+$ spectra reported by Velegrakis and coworkers featured local abundance maxima at $n = 4$ and 6, reminiscent of the anomalies that we observe for $Kr_nAg^+$ and $Xe_nAg^+$. For $Ar_nCu^+$ the relative size is 0.61, even smaller than that of $Kr_nAg^+$ and $Xe_nAg^+$. DFT calculations do, in fact, show abrupt decreases in the dissociation energies beyond $Ar_4Cu^+$ which has $D_{2h}$ rhombic symmetry with $Cu^+$ at its center, and $Ar_6Cu^+$ which forms a distorted octahedron and closes the first solvation shell.

We have previously reported mass spectra of $He_nCu^+$ clusters, formed by doping neutral HNDs with copper and subsequent electron ionization.[83] For this system the relative size is 0.77, slightly larger than that of $Ne_nCu^+$. Local maxima in the ion abundance were observed at $n = 6$, 12, and 24. The anomaly at $n = 12$ was tentatively assigned to an icosahedral solvation shell.

Guided by the comparison between Cu-based and Ag-based cluster ions it is tempting to conclude that an icosahedral $Ng_{12}Ag^+$ represents the first solvation shell for Ne, that $Ng_4Ag^+$ is the largest planar structure for Kr and Xe, and that a distorted octahedral $Ng_6Ag^+$ closes the first solvation shell for Kr and Xe. However, there is another important difference between $Ag^+$, $Cu^+$, and $Au^+$, namely the nature of the bond between the metal ion and the noble gas.[81,84] For gold, the best studied coinage metal, the bond has significant covalent character for Xe and Kr, a slight covalent contribution for Ar, but no covalent character for Ne and He.[17,21-22,85] The covalent character in complexes of Ng with $Cu^+$ is weaker compared to $Au^+$. It is even weaker for $Ag^+$ but $XeAg^+$ seems to have some covalent character as well.[23,26-28,84]

Recently we reported mass spectra of $Au^+$ solvated in He, Ne, Ar, Kr, and Xe.[37] For He and Ne, the abundance distributions of $Ng_nAu^+$ featured anomalies at $n = 12$, similar to the ones reported here for $Ng_nAg^+$. In fact, for these lighter noble gases the gold-based and silver-based systems should adopt similar structures because, first, the nature of the metal-ligand bond is supposedly the same, void of any covalent contributions. Second, their relative sizes are identical within 1 %. However, the distribution of $Ne_nAg^+$ featured another anomaly at $n = 2$ which was also observed for the heavier noble gases which are predicted to form strongly bound $Ng_2Au^+$ because of covalent contributions to their bond.[17,20,22,27-28,86-87] The observation of a magic $Ne_2Au^+$ prompted us to question the notion that bonding in $Ne_nAu^+$ has no covalent contribution. On the other hand, the absence of an abundance anomaly at $Ng_2Ag^+$ in the present work is in line with the notion that covalent contributions to bonding in $Ng_nAg^+$ are extremely weak or entirely absent.[23,26-28,84]

A word of caution is in order though. Bauschlicher et al. have computed dissociation energies of various metal ions including $Cu^+$ (but not $Ag^+$, $Au^+$) bound with one or two noble gas atoms, usually Ar.[88] For some metals, including $Cu^+$, they found that $D_2$ was significantly larger than $D_1$. The increase could be accounted for by promotion and *sp* or *sd* hybridization on the metal ion which may reduce the repulsion between the ligands, rather than by charge transfer from the metal ion to the ligands.

For $Ar_nAu^+$ we observed abundance anomalies at $n = 2$, 6, 9.[37] Computational work by Zhang et al. at the B3LYP level for $n \leq 6$ showed a strong drop in $D_n$ after $n = 2$, and octahedral symmetry for the first solvation shell at $n = 6$.[87] Our own molecular dynamics simulations, based on two-body ligand and ligand-ligand interaction, reproduced the enhanced ion stability for $n = 6$ and 9. The structure of the magic $Ar_9Au^+$ vaguely resembles that of a capped square antiprism.[37] The abundances of $Kr_nAu^+$ and $Xe_nAu^+$ did not feature significant anomalies beyond the magic $n = 2$.

The ion abundance of $Ar_nAg^+$ suggests enhanced stability at $n = 6$ and 12 (Fig. 2d). It seems to present a transitional case, from icosahedral structure for the lighter noble gases to octahedral structure for the heavier ones. It is challenging to come up with structural models for the observed anomalies. If the first solvation shell were to close at an octahedral $Ar_6Ag^+$ and a second layer would grow by adding Ar to its facets, the next shell would close at $n = 14$, not 12. If, on the other hand, the first shell would close at the icosahedral $Ar_{12}Ag^+$ it is difficult to understand why the correspondingly large distance between ligands in an octahedral $Ar_6Ag^+$ would endow the system with enhanced stability. Ab-initio calculations will be needed to assign structures.

We now turn to our findings for $Ag_2^+$ solvated in noble gases. The ion abundance of $Ng_nAg_2^+$ indicates enhanced stability at $n = 6$ and 12 for Ng = He, and at $n = 12$ for Ng = Kr (Fig. 2). We will focus on the anomaly at $n = 12$; an explanation for the enhanced stability of $He_6Ag_2^+$ will require ab-initio calculations.

The computed bond length of $Ag_2^+$ equals 2.89 Å.[89] This is not small compared to the ion-ligand distance ($R_{M-Ng}$ = 2.41 and 2.68 Å for He and Kr, respectively);[18] a solvated $Ag_2^+$ would significantly distort a nominally icosahedral solvation shell. On the other hand, the ion abundance of pure argon cluster ions indicates enhanced stability for $Ar_{14}^+$;[61] this cluster ion consists of an ionic $Ar_4^+$ core plus 10 Ar atoms.[59] In a simplified, earlier model, $Ar_{14}^+$ consists of an ionic $Ar_2^+$ core surrounded by 12 atoms in an approximately icosahedral arrangement.[90] The bond length of $Ar_2^+$ (2.44 Å [91]) is much smaller than the bond length of neutral $Ar_2$ (3.82 Å); apparently the dimer ion can be accommodated at the center of the cluster which has enhanced stability because 12 atoms in the solvation shell optimize the arrangement.

**$H_2O$ solvated in Helium.** Finally we turn to the abundance distribution of $He_nH_2O^+$ which suggests enhanced stability for $n = 12, 32, 44$ (see Fig. 2h). The same sequence of magic numbers has been observed before for $He_nAr^+$,[41] $(H_2)_nH^-$ and its deuterated form,[42] and $(H_2)_nCs^+$.[43] A structural model, originally proposed for $Ar_nNa^+$,[44] attributes the magic numbers to a set of three concentric solvation shells that contain 12, 20, and 12 ligands with icosahedral ($I_h$) symmetry, i.e. the ligands form an icosahedron, dodecahedron, and icosahedron, respectively. The sequence is rarely observed because the appearance of solid-like order in not just one but three solvation shells requires an exact match between the ion-ligand distance and the ligand-ligand distance. The solvent has to form a snowball,[92] i.e. the interaction of the solvent with the ion has to be sufficiently strong to achieve a solvent density that exceeds the density of the bulk in its solid phase. Theoretical work has demonstrated that the size match is just right for $He_nAr^+$ and $(H_2)_nH^-$.[45-46] Apparently, the match is also just right for $He_nH_2O^+$ even though the non-spherical symmetry of the ionic core is incompatible with the nominally icosahedral symmetry of its three solvation shells.

**Conclusion**

We have reported mass spectra of silver cations solvated in noble gases Ng = He, Ne, Ar, Kr, Xe. A novel approach was explored to form $Ng_nAg_m^+$ ions, namely by doping size-to-charge selected $He_N^{z+}$ nanodroplets with Ag and Ng atoms. For all noble gases, the abundance distributions of $Ng_nAg^+$ exhibit one or more anomalies that suggest enhanced stability. By and large, the trend in these magic numbers conforms with expectations based on a simple hard-sphere model, namely a decrease of the coordination number with increasing size of the ligand. Several features, however, are not explained by the hard-sphere model. For example, $He_nAg_2^+$ and $Ar_nAg^+$ both seem to be particularly stable if $n = 6$ or 12. If the clusters were to adopt icosahedral structure for $n = 12$, what is their structure for $n = 6$? Is the striking anomaly at $Kr_{12}Ag_2^+$ really due to an approximately icosahedral arrangement of the solvation shell? Another intriguing observation is the sequence of anomalies at $n = 12, 32, 44$ for $He_nH_2O^+$. The only other systems that exhibit this sequence of magic numbers, commonly attributed to concentric solvation shells of icosahedral symmetry, are $He_nAr^+$,[41,45] $(H_2)_nCs^+$,[43] and $(H_2)_nH^-$.[42,46] Is it coincidental that all these systems involve ions solvated in a quantum fluid?[93] Hopefully, the experimental results presented here will spur further theoretical work.

Acknowledgement
This work was supported by the Austrian Science Fund, FWF (Projects P23657, I4130, and P31149), and the European Commission (ELEvaTE H2020 Twinning Project, Project No. 692335)

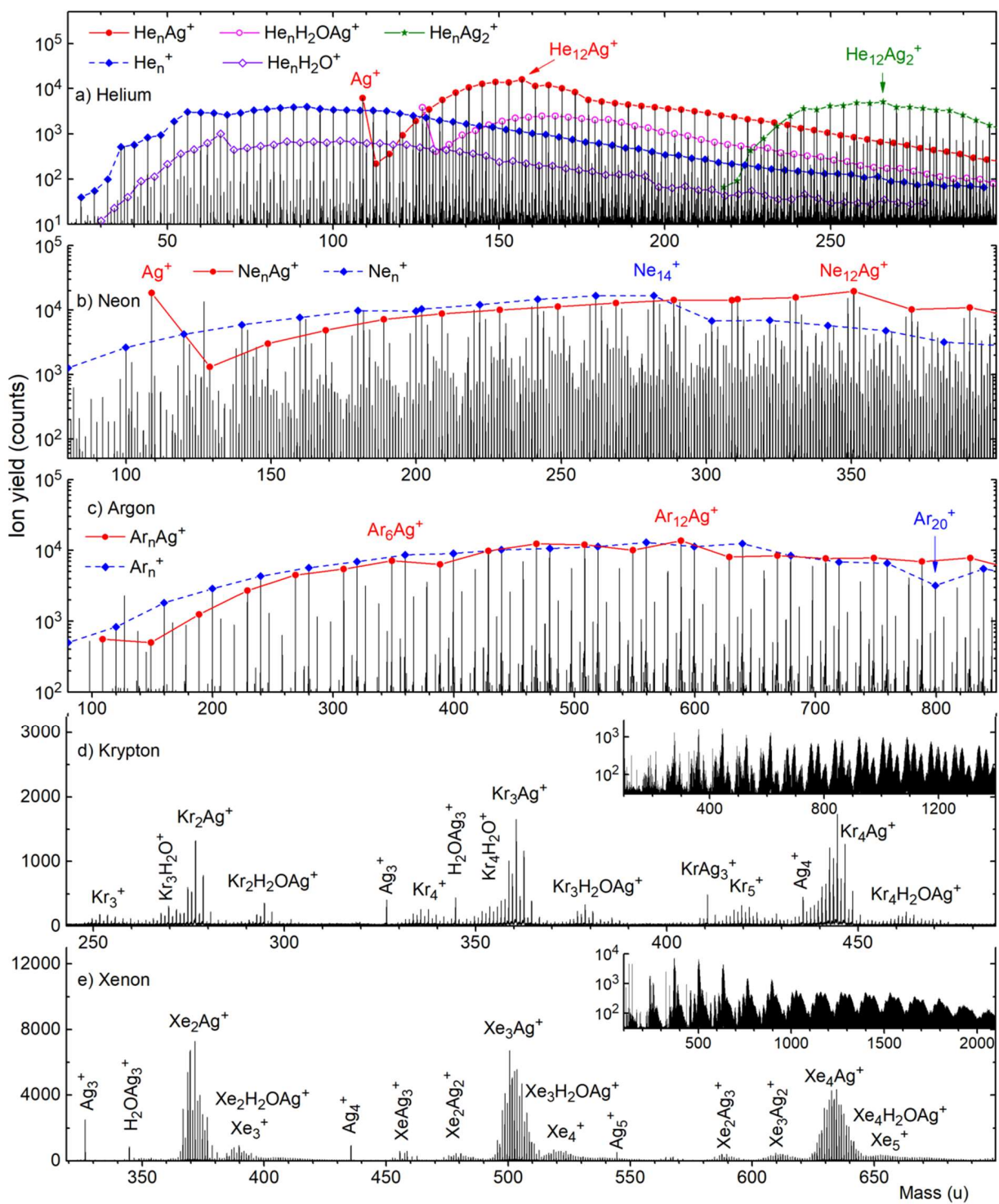

Fig. 1. Sections of mass spectra of helium nanodroplets doped with silver and neon, argon, krypton, xenon (panels b through e, respectively), and a spectrum of undoped nanodroplets (panel a). Insets in panels d and e cover extended mass ranges. In panels a through c, mass peaks due to prominent cluster ion series are marked by symbols and connected by lines.

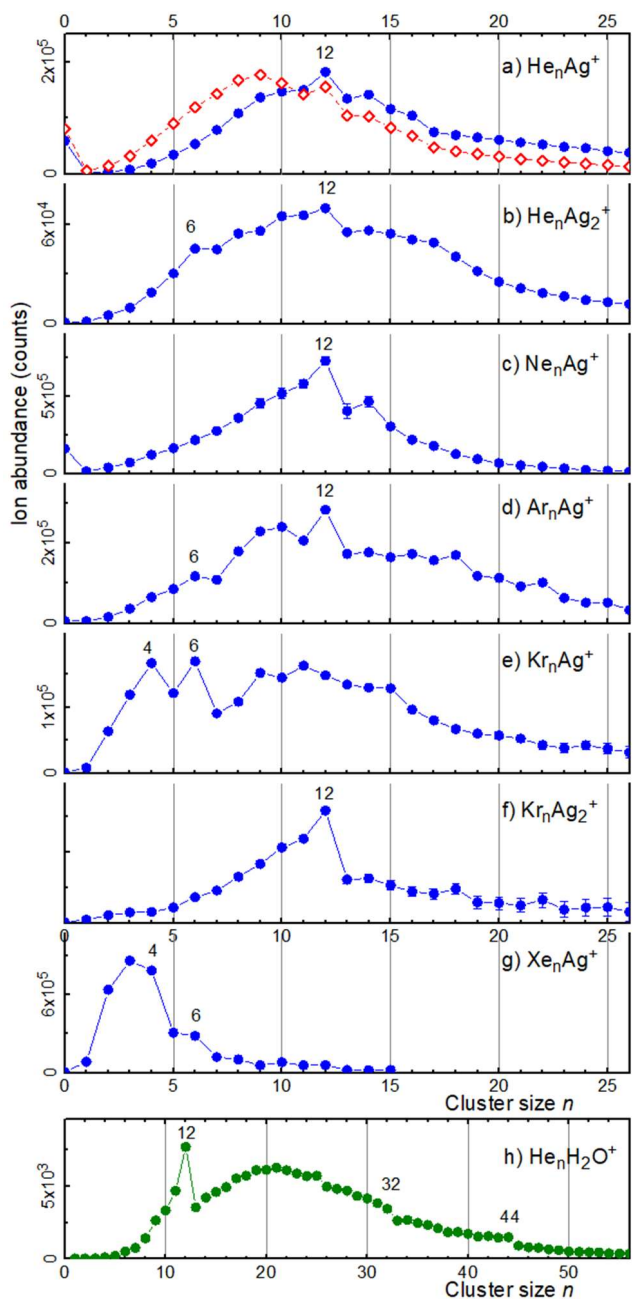

Fig. 2. Abundance of Ng$_n$Ag$^+$ cluster ions (Ng = He, Ne, Ar, Kr, Xe) and a few other ion series. Significant local anomalies in the ion abundances are marked. The two distributions shown in panel a were recorded with different helium pressures in the stripping cell (0.0149 and 0.0168 Pa, respectively). Panel h shows the distribution of He$_n$H$_2$O$^+$; note the expanded size range.

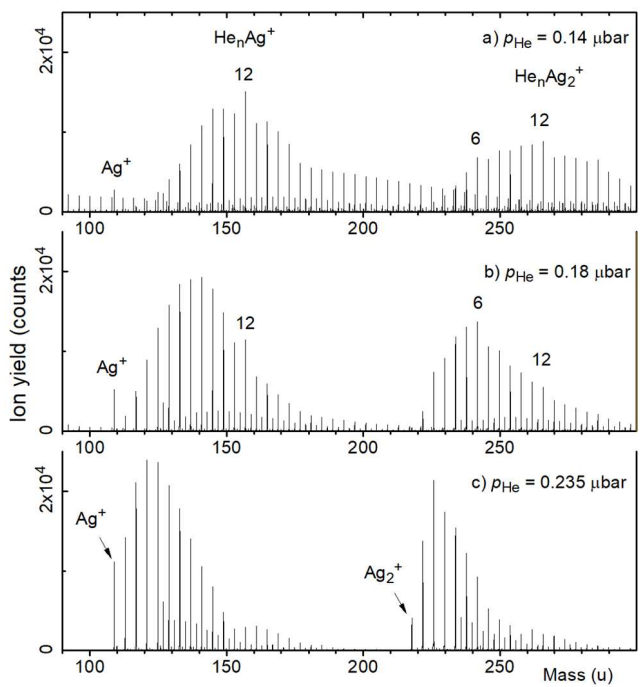

Fig. 3. Mass spectra of charged HNDs doped with silver, recorded with different helium pressures $p_{He}$ in the stripping cell. The distributions of $He_nAg^+$ and $He_nAg_2^+$ shift to toward smaller $n$ as $p_{He}$ increases.

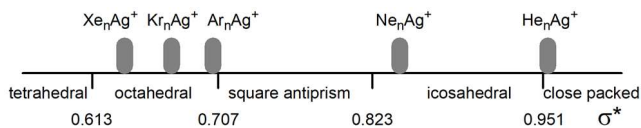

Fig. 4. Data below the abscissa indicate the structure of the first solvation shell for a metal ion ($M^+$) solvated in a noble gas (Ng), predicted within a hard-sphere model as a function of the relative size $\sigma^* = R_{M-Ng}/R_{Ng-Ng}$ where $R_{M-Ng}$ is the distance between the ion and the ligand, and $R_{Ng-Ng}$ is the distance between adjacent ligands.[38] $\sigma^*$ values estimated for $M^+ = Ag^+$ and Ng = He, Ne, Ar, Kr, Xe are indicated above the abscissa.

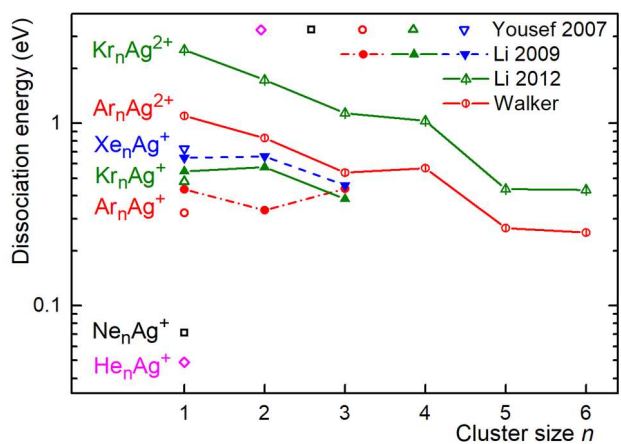

Fig. 5. Published dissociation energies calculated for $Ng_nAg^{z+}$ ions. Open symbols for diatomic $NgAg^+$ are from Yousef et al.;[18] solid symbols connected by lines for $Ng_nAg^+$ ($n \leq 3$; Ng = Ar, Kr, Xe) are from Li et al..[26-28] Also included are energies computed for $Ar_nAg^{2+}$ and $Kr_nAg^{2+}$ dications.[30,48]